\documentclass[11pt,fleqn]{article}
\usepackage{amssymb,latexsym,amsmath,amsfonts,graphicx, ebezier}

    \advance\textheight by \topskip

    \numberwithin{equation}{section}

    \def\tr{{\rm tr \,}}
    \def\Re{{\rm Re \,}}
    \def\Im{{\rm Im \,}}
    \def\Ai{{\rm Ai \,}}

    \def\bigO{{\cal O}}

    \def\P2n{{\rm P}_{{\rm II}}^{(n)}}

    \newtheorem{theorem}{Theorem}[section]

    \newtheorem{Definition}[theorem]{Definition}
    
    \newtheorem{Remark}[theorem]{Remark}
    \newenvironment{remark}{\begin{Remark}\rm}{\end{Remark}}
    \newtheorem{Example}[theorem]{Example}
    \newenvironment{example}{\begin{Example}\rm}{\end{Example}}
    \newtheorem{Assumptions}[theorem]{Assumptions}

    {\rm \trivlist \item[\hskip \labelsep{\bf Proof. }]}%
    {\hspace*{\fill}$\Box$\endtrivlist}
    {\rm \trivlist \item[\hskip \labelsep{\bf Proof}]}%
    {\hspace*{\fill}$\Box$\endtrivlist}

    \newcommand\PVint{\mathop{\setbox0\hbox{$\displaystyle\intop$}%
        \hskip0.2\wd0%
        \vcenter{\hrule width0.6\wd0height0.5pt depth0.5pt}%
        \hskip-0.8\wd0%
        }\mskip-\thinmuskip\intop\nolimits}

\begin{document}
\title{Numerical study of higher order analogues of the Tracy--Widom distribution}
\author{T. Claeys and S. Olver}
\maketitle

\begin{abstract}
We study a family of distributions that arise in critical unitary random matrix
ensembles. They are expressed as Fredholm determinants and describe the limiting
distribution of the largest eigenvalue when the dimension
of the random matrices tends to infinity. The family contains the Tracy--Widom distribution and higher order analogues of
it. We compute the distributions numerically by solving a Riemann--Hilbert problem numerically, plot the distributions, and discuss several properties that they appear to exhibit.
\end{abstract}

\section{Introduction}

Consider the space of Hermitian $n\times n$ matrices with a
probability measure of the form
\begin{equation}\label{random matrix model}
\frac{1}{Z_{n}}e^{-n\,\tr V(M)}dM,
\end{equation}
where $dM$ is the Lebesgue measure defined by
\[dM=\prod_{j=1}^n M_{jj}\prod_{i<j}d\Re M_{ij} d\Im M_{ij},\]
and where the external field $V$ is a real analytic function with
sufficient growth at $\pm\infty$. The quadratic case
$V(M)=\frac{1}{2}M^2$ corresponds to the Gaussian Unitary Ensemble
(GUE) and consists of matrices with independent Gaussian entries.
The probability measure (\ref{random matrix model}) is invariant
under conjugation with a unitary matrix and induces a probability
measure on the matrix eigenvalues given by
\begin{equation}
\frac{1}{\tilde
Z_n}\prod_{i<j}(\lambda_i-\lambda_j)^2\prod_{j=1}^n e^{-nV(\lambda_j)}d\lambda_j.
\end{equation}
The large $n$ limit of the average counting measure of the
eigenvalues of a random matrix (hereafter referred to as the limiting mean eigenvalue
distribution) exists and is characterized as the unique measure
$\mu_V$ which minimizes the logarithmic energy \cite{Deift}
\begin{equation}
I_V(\mu)=\iint \log\frac{1}{|x-y|}d\mu(x)d\mu(y)+\int V(x)d\mu(x),
\end{equation}
among all probability measures $\mu$ on $\mathbb R$. The equilibrium
measure $\mu_V$ depends on the external field $V$ and is supported on a
finite union of intervals $\cup_{j=1}^\ell[a_j,b_j]$, with $\ell, a_j, b_j$
depending on $V$. It is absolutely continuous with a density
$\psi_V$ of the form \cite{DKM}
\begin{equation}
\psi_V(x)=\prod_{j=1}^\ell \sqrt{(b_j-x)(x-a_j)}\ h(x),\qquad x\in
\cup_{j=1}^\ell[a_j,b_j],
\end{equation}
where $h(x)$ is a real analytic function on $\mathbb R$. Generically
\cite{KM}, $h$ has no zeros on the support and in particular at the
endpoints $a_j, b_j$, so that the equilibrium density has square
root vanishing at the endpoints. However, there exist singular
external fields $V$ which are such that $h(b_j)=0$ (or $h(a_j)=0$).
If $h(b_j)=0$, then necessarily an odd number of derivatives vanish:
we have \begin{equation}\label{hb} h'(b_j)=\cdots = h^{(2k-1)}(b_j)=0, \qquad
h^{(2k)}(b_j)\neq 0,\end{equation} for $k\in\mathbb N$. The value $k=0$
corresponds to the generic square-root behavior and $k=1$ to the
first type of critical behavior where $\psi_V(x)\sim c(b_j-x)^{5/2}$
as $x\nearrow b_j$.
\begin{example}
The simplest critical case $k=1$ occurs, e.g., for the critical quartic potential
    \begin{equation}\label{V}
        V(x)=\frac{1}{20}x^4-\frac{4}{15}x^3+\frac{1}{5}x^2+\frac{8}{5}x.
    \end{equation}
    The limiting mean eigenvalue density is then supported on $[-2,2]$ and given by \cite{CIK}
    \begin{equation}
        \psi_V(x) =
        \frac{1}{10\pi}(x+2)^{1/2}(2-x)^{5/2},\qquad\mbox{$x\in [-2,2]$}.
    \end{equation}
\end{example}
\begin{example}
The general situation $k\in\mathbb N$ can be realized for instance with
the following family of polynomials,
\begin{equation}
V_k(x)=2 \cdot \frac{(2k+2)!}{
\left(-\frac{1}{2}\right)_{2k+2}}
\sum_{\ell=0}^{2k+1}\frac{(-1)^\ell(\ell+\frac{1}{2})_{2k+1-\ell}}{(2k+1-\ell)! (\ell+1)}x^{\ell+1},
\end{equation}
where $(a)_m=a\cdot(a+1)\cdots (a+m-1)$ is the Pochhammer symbol.
The corresponding limiting mean eigenvalue densities are given by
\begin{equation}\label{psik}
        \psi_k(x) =
        -\frac{(2k+2)!}{\pi \left(-\frac{1}{2}\right)_{2k+2}}x^{1/2}(1-x)^{2k+1/2},\qquad\mbox{$x\in [0,1]$}.
    \end{equation}
This can be verified by substituting (\ref{psik}) into the
variational conditions that characterize the equilibrium measures
in external field $V_k$. These conditions reduce to the
equation
\[2\PVint_0^1\frac{\psi_k(s)}{x-s}ds=V_k'(x),\qquad\mbox{ for $x\in (0,1)$},\]
which is readily verified.
\end{example}

The random matrix ensembles under consideration are known to have eigenvalues following a determinantal point process with a kernel of the form \cite{Deift}
\begin{equation} \label{kernel}
    K_n(x,y)
        =\frac{e^{-\frac{n}{2}V(x)}
e^{-\frac{n}{2}V(y)}}{x-y}\frac{\kappa_{n-1}}{\kappa_{n}}
(p_n(x)p_{n-1}(y)-p_n(y)p_{n-1}(x)),
\end{equation}
where $p_j$ is the degree $j$ orthonormal polynomial with
respect to the weight $e^{-nV(x)}$ on $\mathbb R$; $\kappa_j$ is its
leading coefficient. Near a right endpoint $b$ of the support for
which $k=0$, it was proved \cite{DKMVZ2, DG} that the large $n$ limit of the
re-scaled kernel is the Airy kernel:
\begin{equation}\label{Airykernel}
\lim_{n\to\infty}\frac{1}{cn^{2/3}}K_n\left(b+\frac{u}{cn^{2/3}},
b+\frac{v}{cn^{2/3}}\right)=K^{(0)}(u,v),
\end{equation}
uniformly for $u,v\geq -L$, $L>0$, where $c=c_V$ and $K^{(0)}$ is
the Airy kernel
\begin{equation}\label{kernel1}
K^{(0)}(u,v)=\frac{\Ai(u)\Ai'(v)-\Ai(v)\Ai'(u)}{u-v}.
\end{equation}
The limit of the probability that the largest eigenvalue is smaller
than $b+\frac{s}{cn^{2/3}}$ is given by a Fredholm determinant:
\begin{equation}\label{TW0}
\lim_{n\to\infty}{\rm
Prob}\left(cn^{2/3}(\lambda_n-b)<s\right)=\det(I-K_s^{(0)}),
\end{equation}
where $K_s^{(0)}$ denotes the Airy-kernel trace-class operator
acting on $L^2(s,+\infty)$. The right hand side of the above
equation is the Tracy--Widom distribution \cite{TW}, which can be written in
the form
\begin{equation}\label{TW}
\det(I-K_s^{(0)})=\exp\left(-\int_s^{+\infty}(y-s)q_0^2(y)dy\right),
\end{equation}
where $q_0$ is the Hastings--McLeod solution $q_0$ of the Painlev\'e
II equation
\begin{equation}\label{PII0}
q_{xx}=xq+2q^3.
\end{equation}
This solution \cite{HM} is characterized by the asymptotic
conditions
\begin{align}&\label{HastingsMcLeod}
q_0(x)\sim \Ai(x), &\mbox{ as $x\to +\infty$,}\\
&\label{HastingsMcLeod2}q_0(x)= \sqrt\frac{-x}{2}\left(1 +
\frac{1}{8x^{3}} + \bigO\left(x^{-6}\right)\right), &\mbox{ as $x\to
-\infty$.}
\end{align}

If $b$ is a right endpoint of the support for which $k=1$, i.e.\ (\ref{hb}) holds with $k=1$ and $b_j=b$, the
 kernel has a different limit. It was showed in
\cite{CV2}, see also \cite{BB, BMP}, that there exists a limiting kernel $K^{(1)}$ such that
\begin{equation}
\lim_{n\to\infty}\frac{1}{cn^{2/7}}K_n\left(b+\frac{u}{cn^{2/7}},
b+\frac{v}{cn^{2/7}}\right)=K^{(1)}(u,v),
\end{equation}
uniformly for $u,v$ in compact subsets of the real line. Since both sides of this equation tend to $0$ rapidly as $u$ or $v$
tends to $+\infty$, this result can be extended to a uniform statement for
$u,v\geq -L$, $L>0$, which implies
\begin{equation}\label{TW1}
\lim_{n\to\infty}{\rm
Prob}\left(cn^{2/7}(\lambda_n-b)<s\right)=\det(I-K_s^{(1)}),
\end{equation}
where $K_s^{(1)}$ is the trace class operator with kernel $K^{(1)}$
on $(s,+\infty)$. In more complicated double scaling limits, where
$V$ is not fixed but depends on $n$ in a critical way, a limiting
kernel $K^{(1)}(u,v;t_0,t_1)$ depending on two parameters was
obtained. Those double scaling limits can describe phase transitions
where the number of intervals in the support of the limiting mean
eigenvalue distributions changes. If we consider an external field
$V=V_t$ depending on a parameter $t$, it can happen that the support
consists of two intervals for $t<1$, but of only one interval for
$t=1$. This happens for example when two intervals approach each
other and simultaneously one of the intervals shrinks in the limit
$t\searrow 1$. Such phase transitions are covered by the kernels
$K^{(1)}(u,v;t_0,t_1)$ with non-zero parameters $t_0, t_1$.

The kernels $K^{(1)}(u,v;t_0,t_1)$ are related to the
second member of the Painlev\'e I hierarchy, but are most easily
characterized in terms of a Riemann-Hilbert (RH) problem. We will
give more details about this in the next section.

Although no proofs have been given for $k>1$, when considering in
more detail the results and methods in \cite{CV2}, one expects a result of the
form
\begin{equation}
\lim_{n\to\infty}\frac{1}{cn^{2/(4k+3)}}K_n\left(b+\frac{u}{cn^{2/(4k+3)}},
b+\frac{v}{cn^{2/(4k+3)}}\right)=K^{(k)}(u,v;t_0,\dots,t_{2k-1}),
\end{equation}
for general $k>1$, where the kernel now depends on $2k$ parameters,
and
\begin{equation}\label{TWk}
\lim_{n\to\infty}{\rm
Prob}\left(cn^{2/(4k+3)}(\lambda_n-b)<s\right)=\det(I-K_s^{(k)}(t_0,\dots,t_{2k-1})).
\end{equation}
The kernels $K^{(k)}$ are related to the Painlev\'e I hierarchy and
will be characterized in the next section in terms of a RH problem. It was proved in
\cite{CIK} that the Fredholm determinant
$\det(I-K_s^{(k)}(t_0,\dots,t_{2k-1}))$ can be expressed explicitly
in terms of a distinguished solution to the equation of order $4k+2$
in the second Painlev\'e hierarchy, and in addition asymptotics for $\det(I-K_s^{(k)}(t_0,\dots,t_{2k-1}))$ as $s\to \pm\infty$ were obtained.
The asymptotics at $+\infty$ can be derived relatively easy from asymptotic properties of the kernel $K^{(k)}$ and are given by
\begin{equation}\label{kernel +infty}
\log\det(I-K_s^{(k)}(t_0,\dots,t_{2k-1}))=\bigO(e^{-cs^{\frac{4k+3}{2}}}),
\qquad \mbox{ as $s\to +\infty$}.
\end{equation}
The asymptotics as $x\to -\infty$ are more subtle and require a detailed analysis of the Fredholm determinants.
In the simplest case $t_0=\cdots=t_{2k-1}=0$, they are given by
\begin{multline}\label{large gap expansion}
\log \det(I-K_s^{(k)})=-\frac{1}{4(4k+3)}\frac{\Gamma(2k +\frac{3}{2})^2}{\Gamma(\frac{3}{2})^2\Gamma(2k+2)^2}|s|^{4k+3}\\
-\frac{2k+1}{8}\log |s|+\chi^{(k)}+\bigO(|s|^{-\frac{4k+3}{2}}),\qquad\mbox{ as $s\to -\infty$,}
\end{multline}
where $\Gamma(x)$ is Euler's $\Gamma$-function. The constant
$\chi^{(k)}$ has no explicit expression, except for $k=0$, where it
was proved in \cite{DIK, BBD} that $\chi^{(0)}=\frac{1}{24}\log
2+\zeta'(-1)$, and $\zeta(s)$ is the Riemann zeta function.

\medskip

The goal of this paper is to set up a numerical scheme for computing the Fredholm determinants
$\det(I-K_s^{(k)}(t_0,\dots,t_{2k-1}))$, which will allow us to draw plots of the
distributions and their densities, to
 verify formulas (\ref{kernel +infty})
and (\ref{large gap expansion}) numerically, to compute numerical
values for the constants $\chi^{(k)}$, and to formulate a number of questions about the
analytic properties of the distributions (monotonicity, inflection points), based on a closer
inspection of the plots. In the next section, we define the kernels
in a precise way using a RH problem. This
RH characterization will also be used for the numerical analysis
which we explain in more detail in Section \ref{section numerics}.
In Section \ref{section plots} finally, we show plots of the
distributions $\det(I-K_s^{(k)})$ and their densities for several
values of $k$ and the parameters $t_0, \ldots, t_{2k-1}$, and we
will formulate a number of open problems.

\section{Riemann--Hilbert characterization of the kernels}
The kernels $K^{(k)}$ have the form
\begin{equation}\label{kernel2}
K^{(k)}(u,v;t_0, \ldots , t_{2k-1})=\frac{\Phi_1^{(2k)}(u)\Phi_2^{(2k)}(v)-\Phi_1^{(2k)}(v)\Phi_2^{(2k)}(u)}{-2\pi i(u-v)},
\end{equation}
where the functions
$\Phi_j^{(2k)}(w)=\Phi_j^{(2k)}(w;t_0,\ldots,t_{2k-1})$ can be characterized in terms of a RH problem.

\subsubsection*{RH problem for $\Phi$}

\begin{figure}[t]
    \begin{center}
    \setlength{\unitlength}{1mm}
    \begin{picture}(95,47)(0,2)
        \put(30,38){\small $\Gamma_2$}
        \put(18,27){\small $\Gamma_3$}
        \put(30,11){\small $\Gamma_4$}
        \put(75,27){\small $\Gamma_1$}
 \put(3,42){\small $\begin{pmatrix}1&0\\1&1\end{pmatrix}$}
        \put(-5,24){\small $\begin{pmatrix}0&1\\-1&0\end{pmatrix}$}
        \put(3,7){\small $\begin{pmatrix}1&0\\1&1\end{pmatrix}$}
        \put(85,24){\small $\begin{pmatrix}1&1\\0&1\end{pmatrix}$}

        \put(50,25){\thicklines\circle*{.9}}
        \put(51,21){\small 0}

        \put(50,25){\line(-2,1){35}} \put(36,32){\thicklines\vector(2,-1){.0001}}
        \put(50,25){\line(-2,-1){35}} \put(36,18){\thicklines\vector(2,1){.0001}}
        \put(50,25){\line(-1,0){40}} \put(30,25){\thicklines\vector(1,0){.0001}}
        \put(50,25){\line(1,0){35}} \put(70,25){\thicklines\vector(1,0){.0001}}
    \end{picture}
    \caption{The jump contour $\Gamma$ and the jump matrices for $\Phi$. }
        \label{figure: gamma}
    \end{center}
\end{figure}
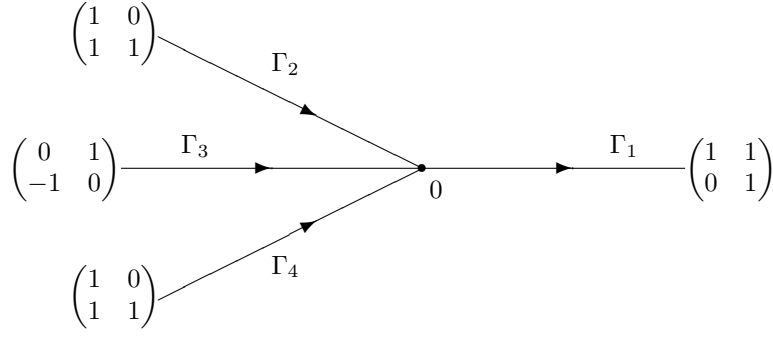

\begin{itemize}
    \item[(a)] $\Phi=\Phi^{(2k)}:\mathbb C\setminus \Gamma\to \mathbb C^{2\times 2}$ is analytic, with \[\Gamma=\cup_{j=1}^4\Gamma_j\cup\{0\}, \qquad \Gamma_1=\mathbb R^+,\quad\Gamma_3=\mathbb R^-, \quad \Gamma_2=e^{\frac{-i\pi}{4k+3}}\mathbb R^-, \quad \Gamma_4=e^{\frac{i\pi}{4k+3}}\mathbb R^-,\] oriented as in Figure \ref{figure: gamma}.
    \item[(b)] $\Phi$ has continuous boundary values $\Phi_+$ as $\zeta$ approaches
$\Gamma\setminus\{0\}$ from the left, and $\Phi_-$,
from the right. They are related by the jump conditions \begin{equation}\label{RHP Phi: b}\Phi_+(\zeta)=\Phi_-(\zeta)S_j,\qquad \mbox{ for $\zeta\in\Gamma_j$,}
\end{equation}
 where
 \begin{align}
        \label{S1}
        &S_1=\begin{pmatrix}
            1 & 1 \\
            0 & 1
        \end{pmatrix},\\
        \label{S24}
        &S_2=S_4=\begin{pmatrix}
            1 & 0 \\
            1 & 1
        \end{pmatrix},\\
        \label{S3}
        &S_3=
        \begin{pmatrix}
            0 & 1 \\
            -1 & 0
        \end{pmatrix}.
    \end{align}
    \item[(c)] $\Phi$ has the following behavior as $\zeta\to\infty$:
    \begin{equation}\label{RHP Psik: c}
        \Phi(\zeta)=\zeta^{-\frac{1}{4}\sigma_3}N\left(I+h\sigma_3\zeta^{-1/2}+
    \bigO(\zeta^{-1})\right)
        e^{-\theta(\zeta)\sigma_3},
    \end{equation}
    where $h=h(t_0, \ldots , t_{2k-1})$ is independent of $\zeta$, $\sigma_3$ is the Pauli matrix $\begin{pmatrix}1&0\\0&-1\end{pmatrix}$, $N$ is given by
    \begin{equation}\label{def: N0}
        N=\frac{1}{\sqrt 2}
        \begin{pmatrix}
             1 & 1 \\
             -1 & 1
        \end{pmatrix}e^{-\frac{1}{4}\pi i\sigma_3},
    \end{equation}
    and
    \begin{equation}\label{def theta0}
        \theta(\zeta;t_0,\ldots,t_{2k-1})=\frac{2}{4k+3}\zeta^{\frac{4k+3}{2}}-2\sum_{j=0}^{2k-1}\frac{
        (-1)^{j}t_{j}}{2j+1}\zeta^{\frac{2j+1}{2}},
    \end{equation}
    where the fractional powers are the principal branches analytic for $\zeta\in \mathbb C\setminus (-\infty,0]$ and positive for $\zeta>0$.
    \item[(d)] $\Phi$ is bounded near $0$.
\end{itemize}
It was proved in \cite{CIK} that this RH problem is uniquely
solvable for any real values of $t_0,\ldots,t_{2k-1}$. The functions
$\Phi_1=\Phi_1^{(2k)}$ and $\Phi_2=\Phi_2^{(2k)}$ appearing in
(\ref{kernel2}) are the analytic extensions of the functions
$\Phi_{11}$ and $\Phi_{21}$ from the sector in between $\Gamma_1$
and $\Gamma_2$ to the entire complex plane. Alternatively they can
be characterized as fundamental solutions to the Lax pair associated
to a special solution to the $2k$-th member of the Painlev\'e I
hierarchy. We will not give details concerning this alternative
description, since the RH characterization is more direct and more
convenient for our purposes.

\begin{remark}
    The description in terms of differential
equations in the PI hierarchy presents the possibility of computing
these distributions using ODE solvers.  However, similar to the Hastings--McLeod solution (see
\cite{PrahoferSpohnKPZ}),  these solutions  are inherently
unstable; hence, applying this approach in
practice would require the use of high precision arithmetic, which is too computationally expensive to be practical.
On the other hand, the representation in terms of a RH problem is
numerically stable, and therefore is reliable.
\end{remark}

\medskip

Not only the kernel $K^{(k)}$ can be described in terms of a RH problem, but also the logarithmic derivative of the Fredholm determinant can be expressed in terms of a RH problem, which shows similarities with the above one, but is nevertheless genuinely different. We have a formula of the form
\begin{equation}\label{identityFredholmX}
\frac{d}{ds}\log\det(I-K_s^{(k)}(t_0,\ldots,
t_{2k-1}))=\frac{1}{2\pi
i}\left.\left(X_s^{-1}(\zeta)X_s'(\zeta)\right)_{21}\right|_{\zeta\searrow
s},
\end{equation}
where $X_s$ is the unique solution to a RH problem, see \cite[Section 2]{CIK}.

This representation provides relative accuracy, whereas the representation as a Fredholm determinant only provides absolute accuracy  \cite{BornemannFredholmDet}.  However, it requires solving a RH problem for each point of evaluation $s$ and numerical indefinite integration to recover the distributions.  Therefore, the expression in terms of a Fredholm determinant is more computationally efficient.

\section{Numerical study of the distributions}\label{section numerics}

    We will compute the higher order Tracy--Widom distributions by calculating $\Phi$ numerically, using the methodology of \cite{SOPainleveII,SORHFramework}.   Consider the following canonical form for a RH  problem:

\subsubsection*{Canonical form for RH problem for $\Psi$}

\begin{itemize}
    \item[(a)] $\Psi: \mathbb C\setminus \overline\Gamma\to \mathbb C^{2\times 2}$ is analytic, where $\overline\Gamma$ is an oriented contour which is
    the closure
    of the set $\Gamma = \Gamma_1 \cup \cdots \cup \Gamma_\ell$ whose connected components can be M\"obius-transformed
    to the unit interval $M_i : \Gamma_i \rightarrow (-1,1)$, with junction points $\Gamma^*=\overline\Gamma\setminus\Gamma$.
    \item[(b)] $\Psi$ has continuous boundary values $\Psi_+$ as $\zeta$ approaches
$\Gamma$ from the left, and $\Psi_-$, from the
right.  For a given function $G$, they
are related by the jump condition
    \begin{equation}\label{canonical jump}
        \Psi_+(\zeta)=\Psi_-(\zeta) G(\zeta).
        \end{equation}
    \item[(c)] As $\zeta\to\infty$, we have $\lim\Psi(\zeta)=I$.
    \item[(d)] $\Psi$ is bounded near $\Gamma^*$.
\end{itemize}

Define the Cauchy transform
    $${\cal C}_{\Gamma} f(\zeta) = {1 \over 2 \pi i} \int_\Gamma {f(t) \over t - \zeta} dt,$$
and denote the limit from the left  (right) for $\zeta \in \Gamma$ by ${\cal C}_\Gamma^+$ (${\cal C}_\Gamma^-$).  We represent $\Psi$ in terms of the Cauchy transform of an unknown function $U$ defined on $\Gamma$:
    $$\Psi(\zeta) = I + {\cal C}_\Gamma U(\zeta).$$
Plugging this into \eqref{canonical jump} we have the linear equation
    \begin{equation}\label{SIE}{\cal C}_\Gamma^+ U - {\cal C}_\Gamma^- U G = G - I.\end{equation}
We solve this equation using a collocation method.

We approximate $U$ by $U_{\bf n}$ for ${\bf n} =   \{n^{\Gamma_1},\dots,n^{\Gamma_\ell}\}$, which is defined  on each component $\Gamma_i$ of the contour in terms of a mapped Chebyshev series:
    $$U_{\bf n}(x) =  \sum_{j = 0}^{n^{\Gamma_i} - 1} U_j^{\Gamma_i} T_j(M_i(x)),\qquad \hbox{ for } x \in \Gamma_i \hbox{ and } i = 1,\ldots,\ell,$$
where $U_j^{\Gamma_i} \in {\Bbb C}^{2 \times 2}$, and $T_j$ is the $j$-th Chebyshev polynomial of the first kind.  The convenience of this basis is that the Cauchy transforms ${\cal C}_{\Gamma_i}[T_j \circ M_i]$ are known in closed form, in terms of hypergeometric functions which can be readily computed numerically \cite{SOHilbertTransform}.

  For each $\zeta \in \Gamma^*$, let $\Omega_1,\ldots,\Omega_L$ be the subset of components in
  $\Gamma$ that have $\zeta$ as an endpoint.  In other words,
  $M_i(\zeta) = p_i$ where $p_i = \pm 1$ for $i=1,\ldots, L$.
    We say that $U$ satisfies the {\it zero sum condition} if
    $$\sum_{i=1}^L p_i U^{\Omega_i}(\zeta) = 0,$$
where $U^{\Omega_i}$ denotes $U$ restricted to $\Omega_i$.
    The boundedness of $\Psi$ implies that $U$ must satisfy the zero sum condition.

Define the mapped Chebyshev points of the first kind:
    $${\bf x}^{\Gamma_i} = M_i^{-1}\begin{pmatrix}-1\cr\cos{\pi\left(1 - {1 \over n^{\Gamma_i}-1}\right)}\cr\vdots\cr\cos{\pi \over n^{\Gamma_i}-1}\cr1\end{pmatrix}$$
and the vector of
unknown Chebyshev coefficients (in ${\Bbb C}^{2 \times 2}$)
    $${\bf U} = \begin{pmatrix}U_0^{\Gamma_1}  \cr \vdots \cr U_{n^{\Gamma_\ell}-1}^{\Gamma_\ell} \end{pmatrix}.$$
Then we can explicitly construct a matrix $C^-$ such that

    $$C^-{\bf U} = \begin{pmatrix} {\cal C}_{\Gamma}^- U_{\bf n}({\bf x}^{\Gamma_1}) \cr \vdots \cr {\cal C}_{\Gamma}^-  U_{\bf n}({\bf x}^{\Gamma_\ell})\end{pmatrix}$$
holds whenever $U_{\bf n}$ satisfies the zero sum condition \cite{SORHFramework}.  To define ${\cal C}_{\Gamma}^- U_{\bf n}({\bf x}^{\Gamma_i})$ at the endpoints, we use
    $${\cal C}_{\Gamma}^- U_{\bf n}(M_{i}^{-1}(\pm 1)) = \lim_{x \rightarrow \pm 1} {\cal C}_{\Gamma}^- U_{\bf n}(M_i^{-1}(x)),$$
which exists when $U_{\bf n}$ satisfies the zero sum condition.

Thus we discretize \eqref{SIE} by
    $${ L}_{\bf n} {\bf U} = (I + C^-) {\bf U} - C^-{\bf U} G_{\bf n} = G_{\bf n} - I$$
where $G_{\bf n} = (G({\bf x}^{\Gamma_1}) , \ldots , G({\bf x}^{\Gamma_\ell}))^\top$  and the multiplication by $G_{\bf n}$ on the right is defined in the obvious way.

    The remarkable fact is that solving this linear system will generically imply that $U_{\bf n}$ satisfies the zero sum condition if $L_{\bf n}$ is nonsingular; if it does not, $L_{\bf n}$ is necessarily not of full rank, and we can replace redundant rows with conditions imposing the zero sum condition \cite{SORHFramework}.  Taking this possibly modified definition of $L_{\bf n}$, we have the following convergence result.

\def\nnnorm#1{ |\mskip-1.5mu\|\,{#1}\,\|\mskip-1.5mu|}

\begin{theorem}\cite{SORHFramework}
    The $L_\infty$ error of the numerical method is bounded by
        $$C_{\bf n} \|L_{\bf n}^{-1}\|_\infty \, \nnnorm{U -  \bar U_{\bf n}},$$
where $C_{\bf n}$ grows logarithmically with $\max {\bf n}$, $\bar U_{\bf n}$ is the polynomial which interpolates $U$ at ${\bf x}^{\Gamma_1},\ldots,{\bf x}^{\Gamma_\ell}$  and
    $$\nnnorm{f} = \| f\|_\infty + \max_{i} \| (M_i^{-1} )' f_i'   \|_\infty.$$
\end{theorem}

In practice, $\|L_{\bf n}^{-1}\|$ appears to grow at most logarithmically with $\max {\bf n}$ whenever a solution to the RH problem exists.  Therefore, if the solution $U$ is smooth, the numerical method will converge spectrally as $\min {\bf n} \rightarrow \infty$, with $\min {\bf n}$ proportional to $\max {\bf n}$.

To apply the numerical method to the RH problem $\Phi$, we need to reduce it to canonical form.   Define $W(\zeta) = \zeta^{-\sigma_3/4} N  e^{-\theta(\zeta)\sigma_3}$, and we use the notation $W_\pm$ to denote the analytic continuation of $W$ above/below its branch cut along $(-\infty,0)$.  We make the following transformation:
    \begin{equation}\label{PsiToPhi}
    \Phi^{(2k)}(\zeta) = \Psi^{(2k)}(\zeta) \begin{cases}
                W(\zeta)            & |\zeta| > 1 \\
                S_4^{-1}            & |\zeta| < 1\hbox{ and $\zeta$ lies between $\Gamma_4$ and $\Gamma_1$} \\
                S_4^{-1} S_1        & |\zeta| < 1\hbox{ and $\zeta$ lies between $\Gamma_1$ and $\Gamma_2$}\\
                S_4^{-1} S_1 S_2^{-1}        & |\zeta| < 1\hbox{ and $\zeta$ lies between $\Gamma_2$ and $\Gamma_3$}\\
                I               & |\zeta| < 1\hbox{ and $\zeta$ lies between $\Gamma_3$ and $\Gamma_4$}
                                      \end{cases}.
                                \end{equation}
Then $\Psi^{(2k)}$ satisfies the RH problem:

\subsubsection*{RH problem for $\Psi^{(2k)}$}

\begin{figure}[t]
\begin{center}
    \setlength{\unitlength}{1.0truemm}
    \begin{picture}(100,75)(0,25)
    \put(50,50){\circle{30}}
    \put(51,57){\thicklines\vector(1,0){.0001}}
    \put(57,50){\line(1,0){40}}
    \put(43.8,53.1){\line(-2,1){40}}
    \put(43.8,46.9){\line(-2,-1){40}}
    \put(75,50){\thicklines\vector(1,0){.0001}}
    \put(28,61){\thicklines\vector(2,-1){.0001}}
    \put(28,39){\thicklines\vector(2,1){.0001}}

    \put(21,65){\small $\tilde\Gamma_2$}
    \put(21,32){\small $\tilde\Gamma_4$}
    \put(37,49){\small $\tilde\Gamma_{42}$}
    \put(50,59){\small $\tilde\Gamma_{21}$}
    \put(50,39){\small $\tilde\Gamma_{14}$}
    \put(78,51){\small $\tilde\Gamma_1$}
    \end{picture}
    \caption{The jump contour $\tilde \Gamma$ and the jump matrices for
    $\Psi$.}
    \label{figure: CanonicalGamma}
\end{center}
\end{figure}
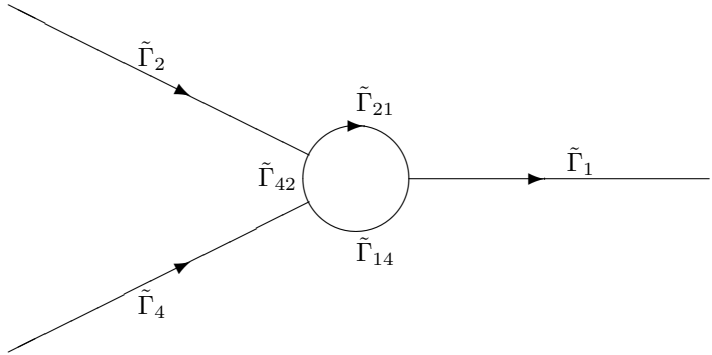

\begin{itemize}
    \item[(a)] $\Psi = \Psi^{(2k)}: \mathbb C\setminus \tilde \Gamma\to \mathbb C^{2\times 2}$ is analytic, with
     \begin{align*}
     \tilde \Gamma &= \tilde\Gamma_1\cup\tilde\Gamma_2\cup \tilde\Gamma_4 \cup \tilde\Gamma_{21} \cup \tilde\Gamma_{42} \cup \tilde\Gamma_{14}\cup\{1,e^{\frac{-i\pi}{4k+3}},e^{\frac{i\pi}{4k+3}}\}, \\
      \tilde \Gamma_1 & = (1,\infty), \quad \tilde \Gamma_2=-e^{\frac{-i\pi}{4k+3}} (\infty,1), \quad \tilde \Gamma_4=-e^{\frac{i\pi}{4k+3}}(\infty,1), \\
      \tilde \Gamma_{21} & = e^{i \pi (1-{1  \over 4 k + 3},0)},  \tilde \Gamma_{42}  = e^{i \pi (-1+{1  \over 4 k + 3},1-{1  \over 4 k + 3})} ,  \tilde \Gamma_{14}  = e^{i \pi (0,-1+{1  \over 4 k + 3})}
      \end{align*}
       oriented as in Figure \ref{figure: CanonicalGamma}.
       \item[(b)] The jump conditions for $\Psi$ are given by
    \begin{align*}
        \Psi_+(\zeta)&=\Psi_-(\zeta)
            W(\zeta) S_j W^{-1}(\zeta),&& \mbox{for $\zeta\in\tilde \Gamma_j$, $j = 1, 2, 4$,}\\
            \Psi_+(\zeta)&=\Psi_-(\zeta)   S_4^{-1}W^{-1}(\zeta) ,&& \mbox{for $\zeta\in\tilde \Gamma_{14}$},\\
              \Psi_+(\zeta)&=\Psi_-(\zeta) S_4^{-1} S_1 W^{-1}(\zeta) ,&& \mbox{for $\zeta\in\tilde \Gamma_{21}$},\\
              \Psi_+(\zeta)&=\Psi_-(\zeta) W_-^{-1}(\zeta) ,&& \mbox{for $\zeta\in\tilde \Gamma_{42}$}.
    \end{align*}
    \item[(c)] As $\zeta\to\infty$, we have $\lim\Psi(\zeta)=I$.
    \item[(d)] $\Psi$ is bounded near $\{1,e^{\frac{-i\pi}{4k+3}},e^{\frac{i\pi}{4k+3}}\}$.
\end{itemize}

With $\Psi$ in this form, we can readily compute it numerically,  recover $\Phi$ by \eqref{PsiToPhi}, and thence evaluate the kernel of  $K_s^{(k)}$ numerically.   This leaves one more task: computing the Fredholm determinant itself.  We   accomplish this using the framework of \cite{BornemannFredholmDet}, which also achieves spectral accuracy.

\begin{figure}
\begin{center}\includegraphics[width=7cm]{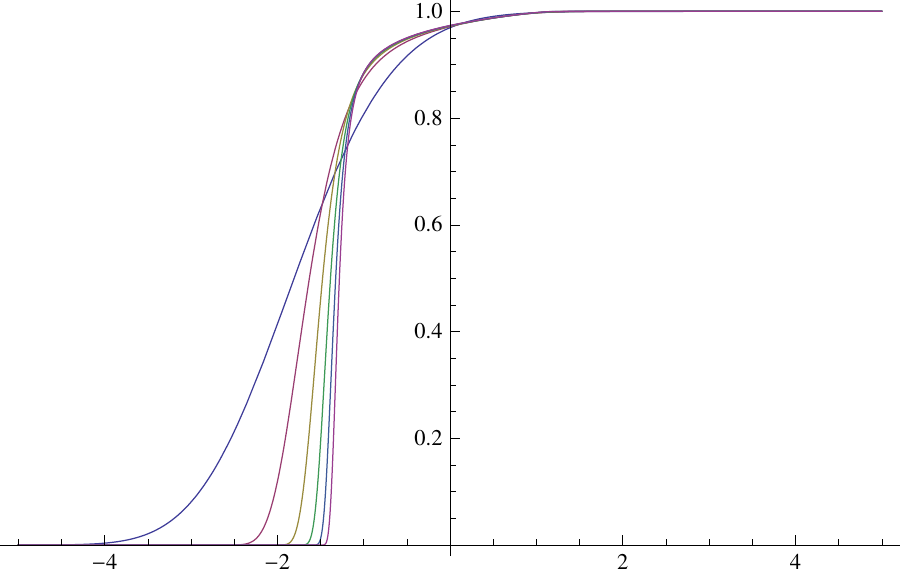}\includegraphics[width=7cm]{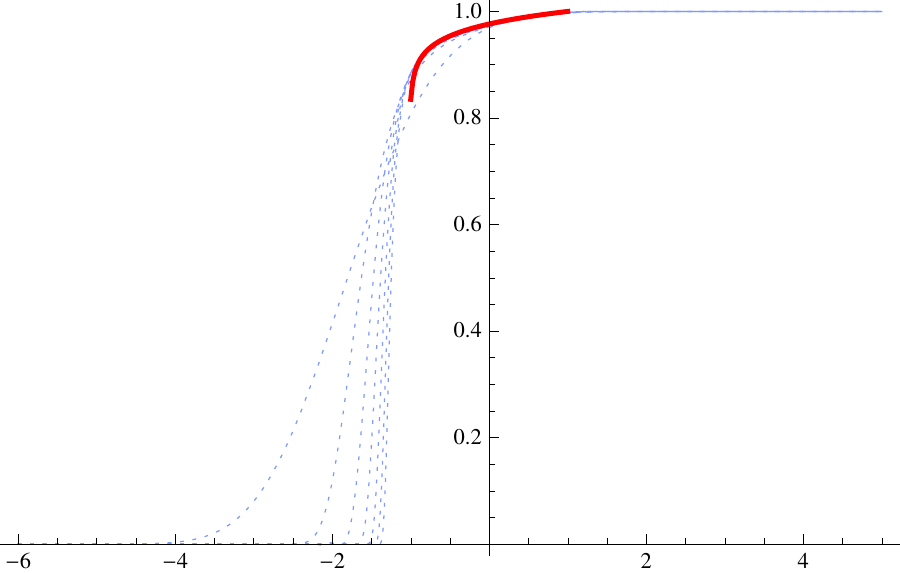}
\caption{The distributions $F_k$ for $k=0,1,\ldots, 5$ with $t_j=0$.
The slope steepens near $-1$ when $k$ increases. On the right, we
 also plot $F_\infty$ (thick curve), as constructed in Section \ref{Largek}.}
        \label{figure: distr}
\end{center}
\end{figure}

\begin{figure}
\begin{center}\includegraphics[width=13cm]{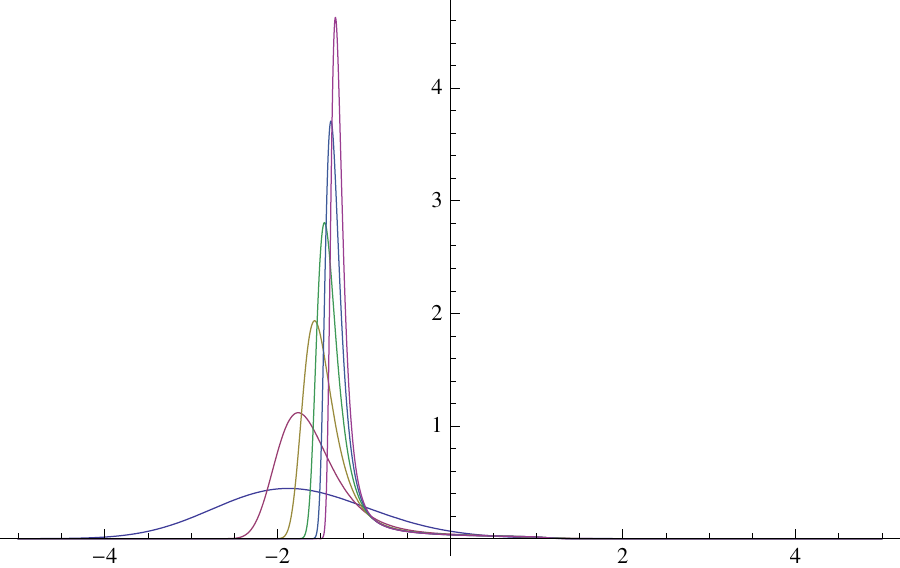} \caption{The densities $F_k'(s)$ for
$k=0,1,\ldots, 5$ with $t_j=0$.}
        \label{figure: dens}\end{center}
\end{figure}

\begin{figure}
\begin{center}\includegraphics[width=7cm]{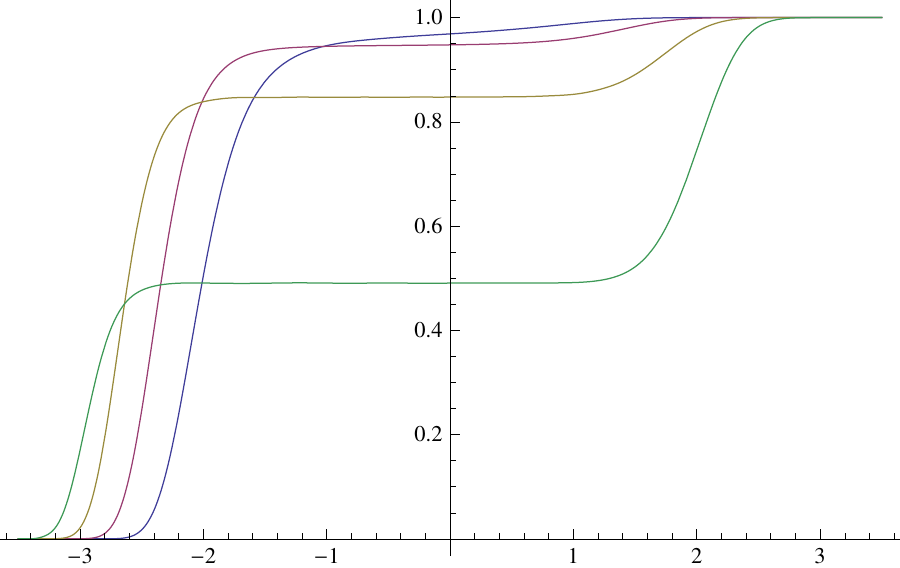}\includegraphics[width=7cm]{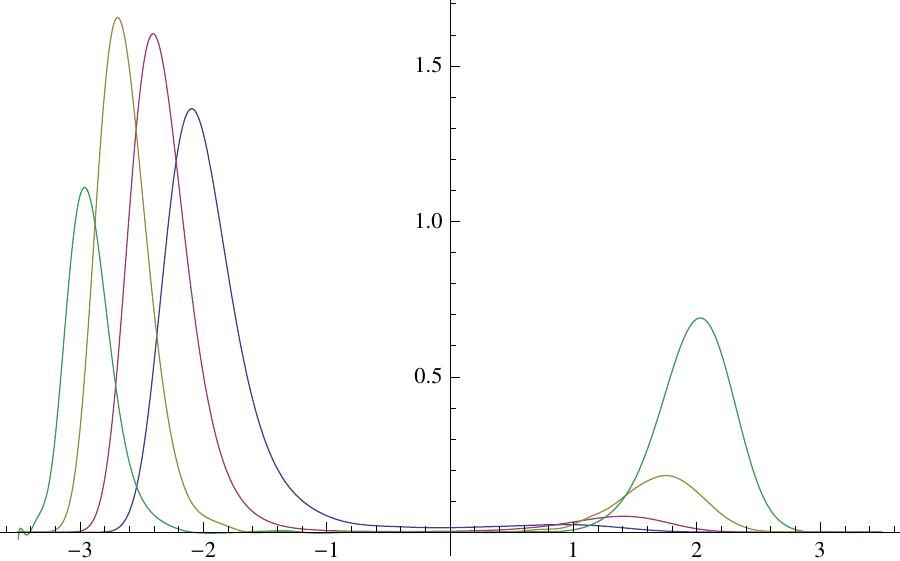}
\caption{The distribution $F_1(s)$ (left) and density $F_1'(s)$ (right) for $t_0=0$ and $t_1=-1,\ldots,-4$. }
        \label{figure: infl}\end{center}
\end{figure}

\section{Plots and open problems}
\label{section plots}
\subsection{Local maxima of the densities}

In Figure \ref{figure: distr}, we plot the numerically computed
distributions $F_k(s;0,\ldots, 0)$ for $k=0,\ldots, 5$, where we
write
\begin{equation}F_k(s;t_0,\ldots, t_{2k-1})=\det(I-K_s^{(k)}(t_0,\ldots, t_{2k-1})).\end{equation}
In Figure \ref{figure: dens} the corresponding densities are drawn.
One observes that each of the densities has only one local maximum
(i.e.\ the distributions have only one inflection point).
The figures suggest that for any $k\in\mathbb N$ and for $t_0=\ldots=t_{2k-1}
= 0$, the densities have only one local maximum.

For general values of the parameters $t_0,\ldots,t_{2k-1}\in\mathbb
R$, the situation is different. We see in Figure \ref{figure: infl},
for $k=1$, $t_0=0$ and varying negative $t_1$, that the densities
have two local maxima. From the random matrix point of view, this
can be explained heuristically by the fact that the kernels
$K^{(1)}(u,v;0,t_1)$ for $t_1<0$ correspond to a double scaling
limit which describes the transition from a random matrix model with a
two-cut support (for the limiting mean eigenvalue distribution) to a
one-cut support, where the parameter $t_1$ regulates the speed of
the transition. To be more precise, consider a random matrix
ensemble with probability measure (\ref{random matrix model}), where
$V=V_n$ depends on $n$. If the dependence of $V$ on $n$ is
fine-tuned in an appropriate way, it can happen that the equilibrium
measure $\mu_{V_n}$ consists of two intervals for finite $n$, but of
only one interval in the limit $n\to\infty$. In order to obtain
$K^{(1)}(u,v;0,t_1)$ as a scaling limit of the eigenvalue
correlation kernel, both intervals in the support of $\mu_{V_n}$
should approach each other and simultaneously one of the intervals
should shrink, as $n\to\infty$. If the $n$-dependence of $V$ is
chosen in an appropriate way, the limiting probability that a random
matrix has an eigenvalue located in the shrinking interval lies
strictly between $0$ and $1$ (it actually increases when $t_1$
decreases). We believe that one local maximum of the densities in
Figure \ref{figure: infl} (the one most to the left) corresponds to
the largest eigenvalue if no eigenvalues lie in the shrinking
interval, and the second local maximum corresponds to the largest
eigenvalue if this one lies in the shrinking interval. For
$k\in\mathbb N$, transitions can take place from at most $k+1$ cuts
to a one-cut regime, and for that reason we expect that for
$k\in\mathbb N$, the density function has at most $k+1$ local
maxima, although we have no analytical evidence for this.

\subsection{Asymptotics as $x\to +\infty$}

\begin{figure}
\begin{center}\includegraphics[width=13cm]{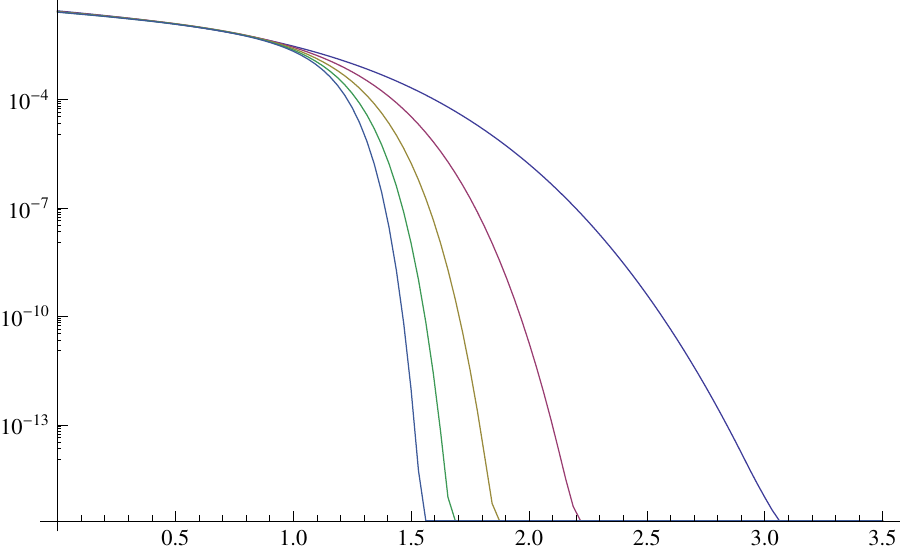}
\caption{$1-F_k(s)$ for $k = 1,\ldots,5$ with $t_j = 0$. }
        \label{figure: positive}\end{center}
\end{figure}

In Figure \ref{figure: positive} we show the rate of convergence to one as $s \rightarrow \infty$ of $F_k(s)$ for various values of $k$.  For $s < 1$, we see that the distribution appears to approach a fixed distribution.  For $s > 1$, the rate of convergence becomes increasingly rapid, matching the asymptotic formula \eqref{kernel +infty}.

\subsection{Asymptotics as $x\to -\infty$}

We note that the constants $\chi^{(k)}$ in (\ref{large gap expansion}) can be expressed as
    \begin{align*}
        \chi^{(k)} &=  \lim_{s \rightarrow - \infty} \left(\log \det(I - K_s^{(k)}) - A_s^{(k)}\right) \\
        & = \lim_{s \rightarrow - \infty} \left(\log \det(I - K_{M}^{(k)}) - \int_{s}^{M}\partial_s \log \det(I - K_{s}^{(k)}) ds -
        A_s^{(k)}\right),
        \end{align*}
with
\[A_s^{(k)}=-\frac{1}{4(4k+3)}\frac{\Gamma(2k +\frac{3}{2})^2}{\Gamma(\frac{3}{2})^2\Gamma(2k+2)^2}|s|^{4k+3}
-\frac{2k+1}{8}\log |s|.\]

For moderate $M$ (we use $M = -.5$), we can reliably calculate $\log
\det(I - K_{M}^{(k)})$ as before. For $s < M$, to reliably calculate
$\partial_s \log \det(I - K_{s}^{(k)})$, we use the RH problem for
$R$ used in \cite[Section 3.5]{CIK}. This RH problem is in canonical
form and $\partial_s \log \det(I - K_{s}^{(k)})$ can be expressed in
terms of its solution. We then expand $\partial_s \log \det(I -
K_{s}^{(k)})-
\partial_s A_s^{(k)}$ in piecewise Chebyshev polynomials, allowing
for the efficient calculation of its integral.

    To verify the accuracy of the above approach, we need to estimate four errors, which we do
    using the following heuristics.  We estimate the error in calculating $\Psi^{(2k)}$ by ensuring that
    the smallest computed Chebyshev coefficient is below a given tolerance ($10^{-12}$).  The error in
    $\log \det(I - K_{M}^{(k)})$ is estimated by examining the Cauchy error as the number of quadrature
    points $m$ in the Fredholm determinant routine increases.  The error in $\partial_s \log \det(I - K_{s}^{(k)})$
    at each point of evaluation $s$ is determined by examining the smallest computed Chebyshev coefficient of the
    numerical approximation to $R$.  Finally, the accuracy of the piecewise Chebyshev approximation to
    $\partial_s \log \det(I - K_{s}^{(k)})$ is estimated by examining each piece's smallest Chebyshev coefficient.

\begin{figure}[t]
\begin{center}\includegraphics[width=7cm]{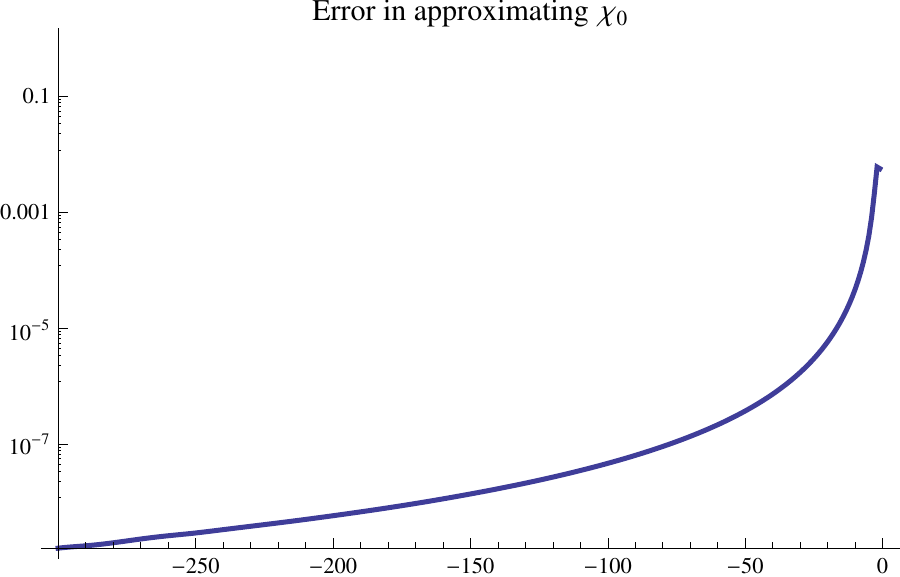}\includegraphics[width=7cm]{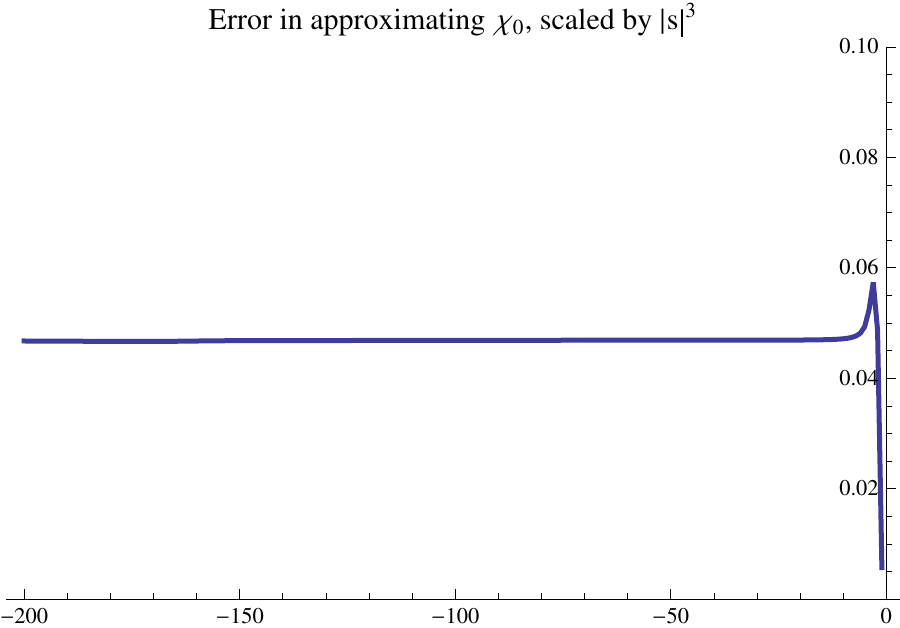}
    \caption{The absolute error in approximating $\chi_0$, $|\chi^{(0)}-\log \det(I - K_{s}^{(0)}) +
      A_s^{(0)}|$ as a function of $s$ (left).    The error multiplied by $|s|^3$ (right), showing faster convergence than predicted.}
    \label{figure: chi0error}
\end{center}
\end{figure}

Using this approach, we estimate the first three $\chi^{(k)}$:
\begin{align*}
    \chi^{(0)} & \approx -0.1365400105,\qquad \hbox{(matches exact expression to 8 digits)}\\
    \chi^{(1)} & \approx -0.09614954 \hbox{ and }\\
    \chi^{(2)} & \approx -0.06145.
\end{align*}
Cancellation and other numerical issues cause the approach to be unreliable for larger $k$.

The convergence to $\chi^{(0)}$ is verified in Figure \ref{figure:
chi0error}.    One interesting thing to note is that the rate of
convergence appears to be faster than predicted:  numerical evidence
suggest convergence like $\bigO(|s|^{-3})$. A similar experiment for
$\chi^{(1)}$ suggests a convergence rate of $\bigO(|s|^{-7})$.  (The
numerics for $\chi^{(2)}$ are insufficiently accurate to make a
prediction.) Therefore, we conjecture that  the error term in
(\ref{large gap expansion}) is in fact $\bigO(|s|^{-(4k + 3)})$,
which is better than the theoretical error $\bigO(|s|^{-{4 k + 3
\over 2}})$.

\subsection{Large $k$ limit}\label{Largek}
For increasing $k$, one observes from Figure \ref{figure: distr} that the slope of the distributions near $-1$ gets steeper.
At first sight, one may expect from Figure \ref{figure: distr} that for large $k$, the distribution function tends to a step function, but a closer inspection reveals that this is not the case. Instead, we believe that there is a limit distribution supported on $[-1,1]$ which is possibly discontinuous at $-1$ but continuous at $1$.

    We present an asymptotic--numerical argument that this is indeed true.  Consider the RH problem for $\Psi^{(2k)}$.
      Note that on $\tilde \Gamma_j$, the jumps $W S_j W^{-1} \rightarrow I$  as $k \rightarrow \infty$.  Furthermore, inside the unit circle $W(\zeta) \rightarrow W^{(\infty)}(\zeta) = \zeta^{-\sigma_3/4} N$.  Finally, $\tilde \Gamma_{42}$
     disappears.  Thus, in a formal sense, we have the following RH problem:

\subsubsection*{RH problem for $\Psi^{(\infty)}$}

\begin{itemize}
    \item[(a)] $\Psi = \Psi^{(\infty)}: \mathbb C\setminus \Gamma\to \mathbb C^{2\times 2}$ is analytic, with
     \begin{align*}
     \Gamma &= \tilde \Gamma_{21} \cup \tilde\Gamma_{14}\cup\{\pm 1\}, \quad  \tilde \Gamma_{21} = e^{i \pi (1,0)},  \tilde \Gamma_{14}  = e^{i \pi (0,-1)}.
      \end{align*}
       \item[(b)] The jump conditions for $\Psi$ are given by (for $W = W^{(\infty)}$)
    \begin{align*}
            \Psi_+(\zeta)&=\Psi_-(\zeta)   S_4^{-1}W^{-1}(\zeta) ,&& \mbox{for $\zeta\in\tilde \Gamma_{14}$},\\
              \Psi_+(\zeta)&=\Psi_-(\zeta) S_4^{-1} S_1 W^{-1}(\zeta) ,&& \mbox{for $\zeta\in\tilde \Gamma_{21}$}.
    \end{align*}
    \item[(c)] As $\zeta\to\infty$, we have $\lim\Psi(\zeta)= I$.
\end{itemize}

This is not in canonical form: the jump matrices are not continuous at $\pm 1$, implying that the solution $\Psi^{(\infty)}$ has singularities.  We rectify this by using local parametrices to remove the jumps.  Define
\begin{equation*}
    P^{(1)}(\zeta)  =\begin{cases}
        \begin{pmatrix}
        1  &0\\
            -1 & 1
        \end{pmatrix}
        \begin{pmatrix}
        1 & {1 \over 2 \pi i} \log\left(-i { z + i \over z - i} - 1\right) \\
           0 & 1
        \end{pmatrix}
        \begin{pmatrix}
        1 &0 \\
            1 & 1
        \end{pmatrix},&\hspace{-1cm}\mbox{$|\zeta-1|<r$, $|z|<1$,}
    \\
 \begin{pmatrix}
        1  &0\\
            -1 & 1
    \end{pmatrix}
     \begin{pmatrix}
        1 & {1 \over 2 \pi i} \log\left(-i { z + i \over z - i} - 1\right) \\
           0 & 1
    \end{pmatrix}
     \begin{pmatrix}
        1 &0 \\
            1 & 1
    \end{pmatrix}S_3S_2W^{-1},\\&\hspace{-1cm}\mbox{$|\zeta-1|<r$, $|z|>1$,}
    \end{cases}
    \end{equation*}
    and
    \begin{equation*}
    P^{(-1)}(\zeta)=
    \begin{cases}
    \begin{pmatrix}
        e^{-2 i \pi/3} & e^{2 i \pi/3}   \\
            1 & 1
    \end{pmatrix} \left(-1 + i{z + i \over z - i}\right)^{\sigma_3/6}
     \begin{pmatrix}
        e^{-2 i \pi/3} & e^{2 i \pi/3}   \\
            1 & 1
    \end{pmatrix}^{-1},\\
     \hspace{8cm}\mbox{ for $|\zeta+1|<r$, $|z|<1$,}\\
    \begin{pmatrix}
        e^{-2 i \pi/3} & e^{2 i \pi/3}   \\
            1 & 1
    \end{pmatrix} \left(-1 + i{z + i \over z - i}\right)^{\sigma_3/6}
     \begin{pmatrix}
        e^{-2 i \pi/3} & e^{2 i \pi/3}   \\
            1 & 1
    \end{pmatrix}^{-1}S_3S_2W_+^{-1},\\ \hspace{8cm}\mbox{ for $|\zeta+1|<r$, $|z|>1$,}\end{cases}
\end{equation*}
with the standard branch cuts, so that they lie on the half circle
$e^{(0,-i \pi)}$.
It is straightforward to verify that $P^{(\pm 1)}$ have the same jumps as $\Psi^{(\infty)}$ inside the disks $|\zeta\mp 1|<r$.

We now define for $r$ sufficiently small,
    $$Y(\zeta) = \begin{cases}\Psi^{(\infty)}(\zeta)P^{(1)}(\zeta)^{-1}, & |\zeta - 1| < r, \\
            \Psi^{(\infty)}(\zeta)P^{(-1)}(\zeta)^{-1}, & |\zeta + 1| < r, \\
            \Psi^{(\infty)}(\zeta), & \mbox{otherwise.}
    \end{cases}$$
Then, $Y$ satisfies a RH problem in canonical form:
\subsubsection*{RH problem for $Y$}

\begin{itemize}
    \item[(a)] $Y : \mathbb C\setminus \overline\Delta\to \mathbb C^{2\times 2}$ is analytic, with
     \begin{align*}
     &\Delta = \Delta_{1} \cup \Delta_{2}\cup \Gamma_r(\pm 1)\cup\{\pm e^{\pm i \theta}\}, \\
     &\Delta_{1} = e^{i \pi (1-\theta,\theta)},\quad   \Delta_{2}  = e^{i \pi (-\theta,\theta-1)},\quad
     \Gamma_r(a)=\{\zeta: |\zeta-a|=r\},
      \end{align*}
      where $\theta$ is given by $r=|e^{i\theta }-1|$.
       \item[(b)] The jump conditions for $Y$ are given by
    \begin{align*}
            Y_+(\zeta)&=Y_-(\zeta)   S_4^{-1}S_1W^{-1}(\zeta) ,&& \mbox{for $\zeta\in \Delta_{1}$},\\
        Y_+(\zeta)&=Y_-(\zeta) S_4^{-1}  W^{-1}(\zeta) ,&& \mbox{for $\zeta\in \Delta_{2}$},\\
        Y_+(\zeta)&=Y_-(\zeta) P^{(\pm 1)}(\zeta) ,&& \mbox{for $\zeta\in \Gamma_r(\pm 1)$}.
    \end{align*}
    \item[(c)] As $\zeta\to\infty$, we have $\lim Y(\zeta)= I$.
    \item[(d)] $Y$ is bounded near $\{\pm e^{\pm i \theta}\}$.
\end{itemize}

We can  compute $Y$, and hence $\Psi^{(\infty)}$ numerically.   We therefore define
    \begin{equation*}
    \Phi^{(\infty)}(\zeta) = \Psi^{(\infty)}(\zeta) \begin{cases}
                0               & |\zeta| > 1 \\
                S_4^{-1}            & |\zeta| < 1\hbox{ and  $\Im \zeta< 0$ }   \\
                S_4^{-1} S_1        & |\zeta| < 1\hbox{ and  $\Im \zeta > 0$ }
                                      \end{cases}.
                                \end{equation*}
which we use to compute the kernel of  $K_s^{(\infty)}$, which is a trace-class operator now acting on $L^2(s,1)$.
(Again, we do not have a rigorous reason why the limiting operator acts only on $L^2(s,1)$, not $L^2(s,\infty)$.
Instead, we justify this  by the accuracy of the numerics.)

It should be noted that the local parametrices $P^{(\pm 1)}$ are not
close to the identity matrix on $|\zeta\mp 1|=r$, and therefore they
would not be suitable parametrices to be used for a rigorous
Deift/Zhou steepest descent analysis \cite{DZ} applied to the RH problem for
$\Psi$. However, this is not an issue here: numerically it is
sufficient that the local parametrices satisfy the required jump
conditions.

    While this construction has not been mathematically justified, it is perfectly usable in a numerical way.
    In fact, the resulting distribution matches the asymptotics for the finite $k$ distributions, cf. Figure
    \ref{figure: distr}, providing strong evidence  that, for $-1 < s < 1$,
      $$\lim_{k \rightarrow \infty} \det(I - K_s^{(k)}) = \det(I - K_s^{(\infty)}).$$
We remark that the $\Phi^{(\infty)}$ appears to be smooth
near $+1$, hence Bornemann's numerical Fredholm determinant routine remains accurate for $s > -1$.  However, the singularity in $\Phi^{(\infty)}$ at $-1$ causes the accuracy to break down as $s$ approaches $-1$.  Therefore, we cannot infer whether the distribution approaches zero smoothly, or if there is a jump.

\section*{Acknowledgements}
TC acknowledges support by the Belgian Interuniversity
Attraction Pole P06/02 and by the ERC program FroM-PDE.

\obeylines \texttt{Tom Claeys, tom.claeys@uclouvain.be, Universit\'e
Catholique de Louvain, Chemin du cyclotron 2, B-1348
Louvain-La-Neuve, BELGIUM
\medskip
Sheehan Olver, olver@maths.usyd.edu.au, School of Mathematics and
Statistics, The University of Sydney, NSW 2006 Australia }


\begin{thebibliography}{99}
\bibitem{BBD}
    J. Baik, R. Buckingham, and J. Di Franco, Asymptotics of Tracy--Widom distributions and the total integral of a Painlev\'e
    II function, {\em Comm. Math.
    Phys.} {\bf 280} (2008), 463--497.
\bibitem{BornemannFredholmDet}
    F. Bornemann, On the numerical evaluation of Fredholm determinants, {\em Math. Comp} {\bf 79} (2010), 871--915.
\bibitem{BB}
        M.J. Bowick and E. Br\'ezin,
        Universal scaling of the tail of the density of
            eigenvalues in random matrix models,
        {\em Phys. Lett. B} {\bf 268} (1991), no. 1, 21--28.
\bibitem{BMP}
        E. Br\'ezin, E. Marinari, and G. Parisi,
        A non-perturbative ambiguity free solution of a string model,
        {\em Phys. Lett. B} {\bf 242} (1990), no. 1, 35--38.
\bibitem{CIK}T. Claeys, A. Its, and I. Krasovsky, Higher order analogues of the Tracy--Widom
distribution and the Painlev\'e II hierarchy, {\em Comm. Pure Appl.
Math.} {\bf 63} (2010), 362--412.
\bibitem{CV2}T. Claeys and M. Vanlessen, Universality of a double
    scaling limit near singular edge points in random matrix
    models, {\em Comm. Math. Phys.} {\bf 273} (2007), 499--532 .
\bibitem{Deift}
    P. Deift,
    ``\,Orthogonal Polynomials and Random Matrices: A  Riemann--Hilbert Approach",
    Courant Lecture Notes 3, New York University 1999.
\bibitem{DG}
    P. Deift and D. Gioev,
    Universality in random matrix
    theory for the orthogonal and symplectic ensembles,
    {\em Int. Math. Res. Pap. IMRP} {\bf 2007} (2007), no. 2, Art. ID rpm004, 116 pp.
    \bibitem{DIK}
P. Deift, A. Its, and I. Krasovsky, Asymptotics for the Airy-kernel determinant,
{\em Comm. Math. Phys.} {\bf 278} (2008), 643--678.
\bibitem{DKM}
    P. Deift, T. Kriecherbauer, and K.T--R McLaughlin,
    New results on the equilibrium measure for logarithmic potentials
    in the presence of an external field,
    {\em J. Approx. Theory} {\bf 95} (1998), 388--475.
\bibitem{DKMVZ2}
    P. Deift, T. Kriecherbauer, K.T--R McLaughlin, S. Venakides, and X. Zhou,
    Uniform asymptotics for polynomials orthogonal with respect to
    varying exponential weights and applications to universality
    questions in random matrix theory,
    {\em Comm. Pure Appl. Math.} {\bf 52} (1999), 1335--1425.
\bibitem{DZ}
        P. Deift and X. Zhou,
        A steepest descent method for oscillatory Riemann--Hilbert problems.
            Asymptotics for the MKdV equation,
        {\em Ann. Math.} {\bf 137} (1993), no. 2, 295--368.
\bibitem{HM}
    S.P. Hastings and J.B. McLeod,
    A boundary value problem associated with the second Painlev\'e transcendent
    and the Korteweg--de Vries equation,
    {\em Arch. Rational Mech. Anal.} {\bf 73} (1980), 31--51.
\bibitem{KM}
    A.B.J. Kuijlaars and K.T--R McLaughlin,
    Generic behavior of the density of states in random matrix theory
    and equilibrium problems in the presence of real analytic external
    fields,
    {\em Comm. Pure Appl. Math.} {\bf 53} (2000), 736--785.

\bibitem{SORHFramework}
    S. Olver,
    A general framework for solving Riemann--Hilbert problems numerically,
    Report no. NA-10/5, Mathematical Institute, Oxford University.

\bibitem{SOPainleveII}
    S. Olver,
    Numerical solution of Riemann--Hilbert problems: Painlev\'e II,
    {\em Found. Comput. Maths} {\bf 11} (2011), 153--179.

\bibitem{SOHilbertTransform}
    S. Olver,
     Computing the Hilbert transform and its inverse,
    {\em Maths Comp.} {\bf 80} (2011), 1745--1767.

\bibitem{PrahoferSpohnKPZ}
    M. Pr{\"a}hofer, and H. Spohn,
    Exact scaling functions for one-dimensional stationary KPZ growth,
    {\em J. Stat. Phys.}  {\bf 115} (2004), 255--279.


\bibitem{TW}
        C.A. Tracy and H. Widom,
        Level spacing distributions and the Airy kernel,
        {\em Comm. Math. Phys.} {\bf 159} (1994), no. 1, 151--174.

\end{thebibliography}
\end{document}